\documentclass[final,1p,times,twocolumn]{elsarticle}
\usepackage{graphicx,amssymb}
\journal{Physics Letters B}
\begin{document}
\begin{frontmatter}

\title{Accurate Calibration of the Velocity-dependent One-scale Model for Domain Walls}
\author[inst1,inst2]{A. M. M. Leite}\ead{up080322016@alunos.fc.up.pt}
\author[inst1]{C. J. A. P. Martins\corref{cor1}}\ead{Carlos.Martins@astro.up.pt}
\author[inst3]{E. P. S. Shellard}\ead{E.P.S.Shellard@damtp.cam.ac.uk}
\address[inst1]{Centro de Astrof\'{\i}sica, Universidade do Porto, Rua das Estrelas, 4150-762 Porto, Portugal}
\address[inst2]{\'Ecole Polytechnique, 91128 Palaiseau Cedex, France}
\address[inst3]{Department of Applied Mathematics and Theoretical Physics, Centre for Mathematical Sciences, University of Cambridge, Wilberforce Road, Cambridge CB3 0WA, United Kingdom}
\cortext[cor1]{Corresponding author}

\begin{abstract}
We study the asymptotic scaling properties of standard domain wall networks in several cosmological epochs. We carry out the largest field theory simulations achieved to date, with simulation boxes of size $2048^3$, and confirm that a scale-invariant evolution of the network is indeed the attractor solution. The simulations are also used to obtain an accurate calibration for the velocity-dependent one-scale model for domain walls: we numerically determine the two free model parameters to have the values $c_w=0.34\pm0.16$ and $k_w=0.98\pm0.07$, which are higher precision than (but in agreement with) earlier estimates.
\end{abstract}

\begin{keyword}
Cosmology \sep Topological Defects \sep Domain Walls \sep Numerical Simulation \sep VOS model
\end{keyword}

\end{frontmatter}

\section{Introduction}
\label{intr}

A key consequence of cosmological phase transitions is the formation of topological defects \cite{Kibble,Book}. While cosmic strings have attracted most of the community's attention, domain walls are useful as a testbed case with which one can gather information relevant for other more complex defects (despite being tightly constrained by observations \cite{Zeldovich,Junctions}). Here we take advantage of ever-improving computing resources to carry out a large set of $2048^3$ high-resolution simulations of domain walls, using the standard Press-Ryden-Spergel (PRS) algorithm \cite{Press}. This is a follow-up on \cite{Leite}, where the results of simulations of size up to $1024^3$ were presented, and we confirm and expand their results.

Early generations of domain wall simulations \cite{Press,Coulson,Larsson,Fossils,Garagounis,MY1,MY2,Lalak} found some hints for late-time deviations from the scale-invariant evolution, which would be the expected behavior \cite{Hindmarsh,MY2}. Our previous work \cite{Leite} found no such deviations, which provided support for the hypothesis that the earlier results were simply a consequence of the limited dynamical range of numerical simulations. We believe that the present work clearly confirms this.

Macroscopic properties of defect networks can be accurately described by an analytic velocity-dependent model, first derived for cosmic strings \cite{ms1a,ms1b,extend}. The large-scale features of the network are described by a characteristic scale $L$ (which one can interchangeably think of as a typical defect separation or correlation length) and a microscopically averaged (root-mean-squared) velocity $v$. This has the advantages of tractability and conceptual simplicity but must include phenomenological parameters which parametrize our ignorance about certain dynamical mechanisms.  The only way to accurately determine the correct values of these parameters is by employing large-scale numerical simulations to calibrate them. The main goal of the current work is precisely to improve this calibration.

\section{Numerical simulations}
\label{prs}

We will study simple (single-field) domain wall networks in flat homogeneous and isotropic Friedmann-Robertson-Walker (FRW) universes. (Throughout the paper we shall use fundamental units, in which $c=\hbar=1$.) A scalar field $\phi$ with Lagrangian density
\begin{equation}
\mathcal{L}={\frac{1}{2}}\phi_{,\alpha}\phi^{,\alpha}-V_{0}\left({\frac{\phi^{2}}{\phi_{0}^{2}}}-1\right)^{2}\,,
\label{potential}
\end{equation}
provides the simplest case. By standard variational methods we obtain the field equation of motion (written in terms of physical time $t$) 
\begin{equation}
{\frac{{\partial^{2}\phi}}{\partial 
t^{2}}}+3H{\frac{{\partial\phi}}{\partial 
t}}-\nabla^{2}\phi=-{\frac{{\partial 
V}}{\partial\phi}}\,.\label{dynamics}
\end{equation}
where $\nabla$ is the Laplacian in physical coordinates, $H=a^{-1}(da/dt)$ is the Hubble parameter and $a$ is the scale factor, which we generically assume to vary as $a\propto t^\lambda$. In what follows we will study the network's evolution in several such cosmological epochs.

We follow the procedure of Press, Ryden and Spergel \cite{Press}, modifying the equations of motion in such a way that the thickness of the domain walls is fixed in co-moving coordinates. The reliability of this method has been numerically tested in previous work \cite{Press,MY1,Moore}. In the PRS method, equation (\ref{dynamics}) becomes: 
\begin{equation}
{\frac{{\partial^{2}\phi}}{\partial\eta^{2}}}+\alpha\left(\frac{d\ln 
a}{d\ln\eta}\right){\frac{{\partial\phi}}{\partial\eta}}-{\nabla}^{2}\phi=
-a^{\beta}{\frac{{\partial 
V}}{\partial\phi}}\,.\label{dynamics2}
\end{equation}
where $\eta$ is the conformal time and $\alpha$ and $\beta$ are constants: $\beta=0$ is used in order to have constant co-moving thickness and $\alpha=3$ is chosen to require that the momentum conservation law of the wall evolution in an expanding universe is maintained \cite{Press}. The specific parameters used in the simulations are $\phi_{0}=1$, $V_0=\pi^2/2W_0^2$, where $W_0=10$ is the wall thickness; these choices are also justified by previous work on this algorithm \cite{Press,MY1,Moore}.

\begin{figure}
\begin{center}
\includegraphics[width=2.6in]{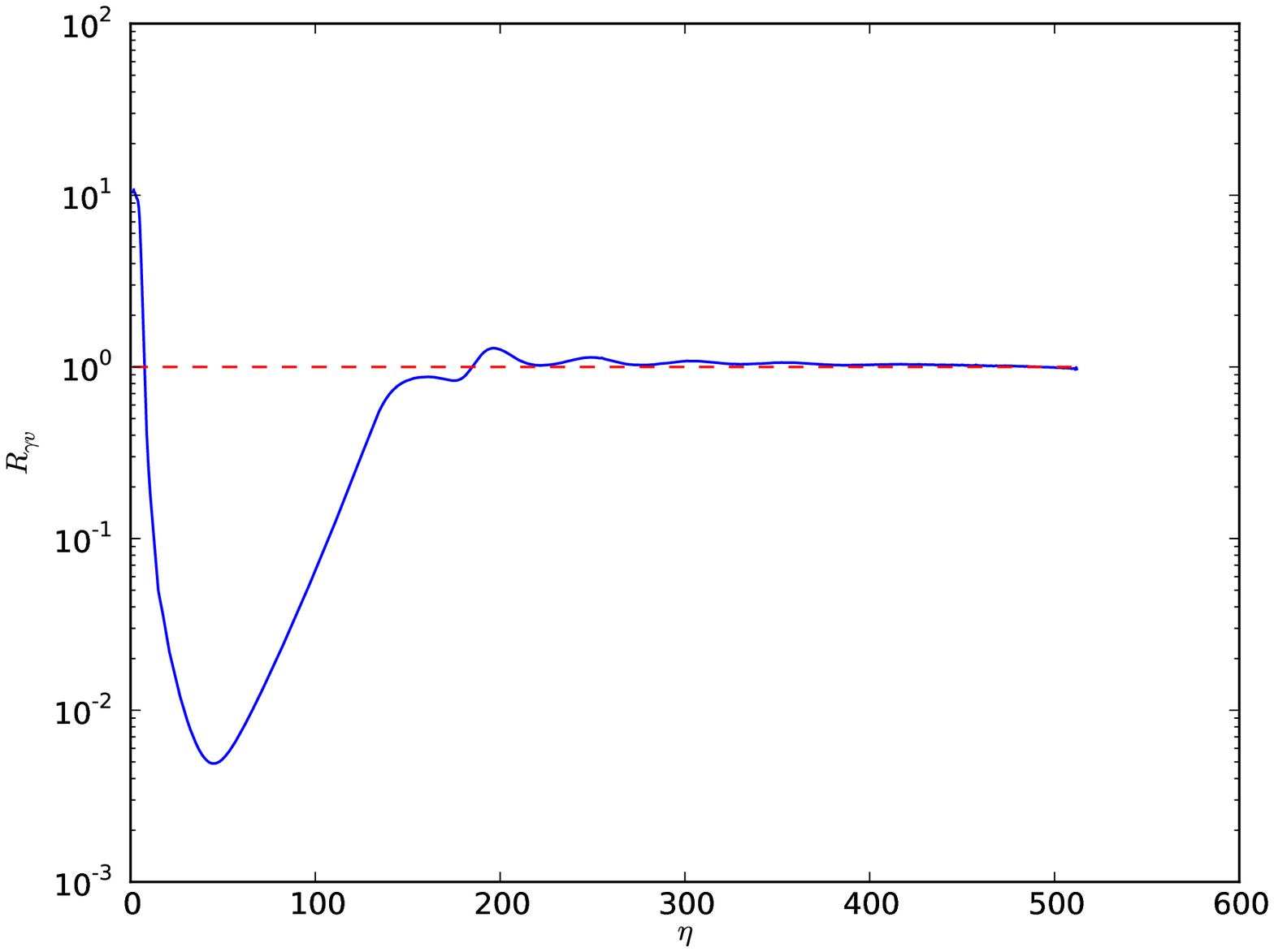}
\includegraphics[width=2.6in]{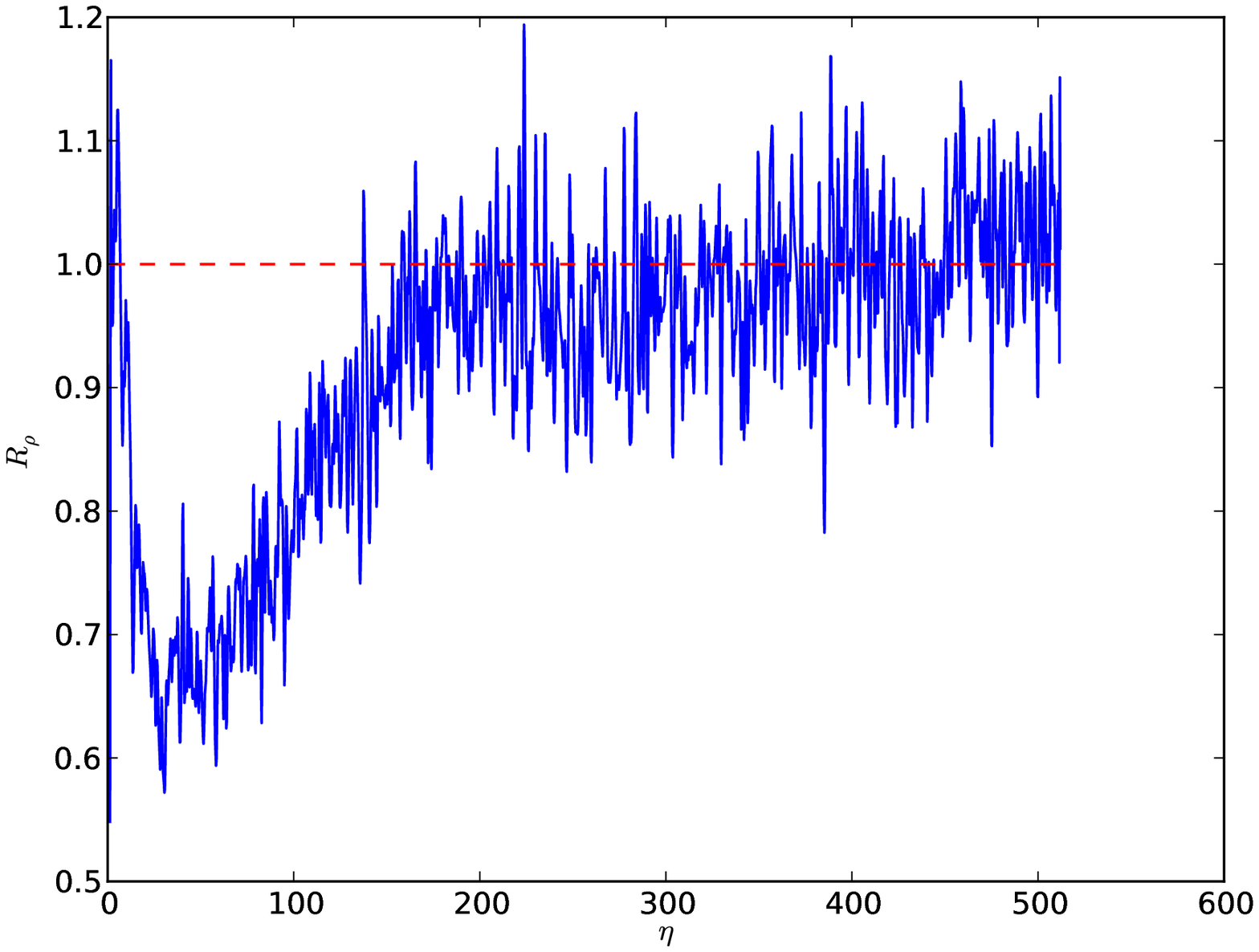}
\end{center}
\caption{\label{fig1}The evolution of the velocity and density ratios ($R_{\gamma v}$ and $R_\rho$, defined in the text) for the average of two sets of ten  $1024^3$ matter era simulations with $W_0=100$ and $W_0=10$. As expected, both ratios become unity after a transient period.}
\end{figure}

Despite these previous results one may wonder whether the chosen wall thickness is sufficient to accurately calibrate the model. Figure \ref{fig1} compares two series of ten $1024^3$ matter era runs with $W_0=10$ and $W_0=100$. Plotted are the ratios of the network velocities ($R_{\gamma v}=(\gamma v)_{W0=100}/(\gamma _v)_{W0=10}$) and densities ($R_{\rho}=(\rho)_{W0=100}/(\rho)_{W0=10}$). After transients due to the choices of initial conditions, both of these ratios converge to unity, with statistical uncertainties well below $10\%$. This convergence is expected to be stronger wilth larger ensembles and/or larger boxes.

Equation (\ref{dynamics2}) is integrated using a standard finite-difference scheme. We assume the initial value of $\phi$ to be a random variable between $-\phi_{0}$ and $+\phi_{0}$ and the initial value of $\partial\phi/\partial\eta$ to be zero. This will lead to large energy gradients in the early timesteps of the simulation, and the network needs some time (which is proportional to the wall thickness) to wash away these initial conditions. Our simulations start at a conformal time $\eta_0=1$ and evolve in timesteps $\Delta\eta=0.25\eta_0$ until a conformal time equal to half the box size (that is, $\eta=1024$).

The conformal time evolution of the co-moving correlation length of the network $\xi_c$ (specifically $A/V\propto \xi_{c}^{-1}$, $A$ being the comoving area of the walls) and the wall velocities (specifically $\gamma v$, where $\gamma$ is the Lorentz factor) are directly measured from the simulations, using techniques previously described in \cite{MY2}. However here we use a newly parallelized version of the code, optimized for the Altix UV1000 architecture of the COSMOS Consortium's supercomputer.

\section{Analytic model}
\label{model}

In order to model a defect network one starts from the microscopic equations of motion (the Nambu-Goto equations, in the case of strings) and, through a suitable averaging, arrives at 'thermodynamic' evolution equations. The non-trivial part of this procedure is the inclusion of terms to account for defect interactions and energy losses. Such terms must be added in a phenomenological way, and for their calibration one must resort to numerical simulations.

For cosmic strings, this procedure leads to the velocity-dependent one-scale (VOS) model \cite{ms1a,ms1b,extend}, which has been thoroughly tested against simulations. One can follow an analogous procedure both for the case of monopoles \cite{Monopoles} and for domain walls. This latter case was first studied in \cite{MY2}, and more recently \cite{Leite} provided a preliminary calibration; here we will provide a more quantitative one.

The evolution equation for the characteristic wall lengthscale $L$ (which is related to the wall density $\rho_w$ via $L=\sigma/\rho_w$, where $\sigma$ is the domain wall energy per unit area) and their RMS velocity $v$, are as follows
\begin{equation}
\frac{dL}{dt}=(1+3v^2)HL+c_wv\,
\label{rhoevoldw1}
\end{equation}
\begin{equation}
\frac{dv}{dt}=(1-v^2)\left(\frac{k_w}{L}-3Hv\right)\,.
\label{vevoldw1}
\end{equation}
Here $c_w$ and $k_w$ are the free parameters: the former quantifies energy losses, while the latter quantifies the (curvature-related) forces acting on the walls. To a first approximation, these are expected to be constant. Note that in the context of the VOS model the characteristic length scale $L$ can further be identified with the physical correlation length $\xi_{phys}$. The comoving version of this was defined in the previous section, and the two are related via
\begin{equation}
\xi_{phys}=a\xi_c\,,
\end{equation}
and we are therefore assuming that $\xi_{phys}\equiv L$. Note that if
\begin{equation}
\xi_c\propto \eta^{1-\delta}\,,
\end{equation}
then, for an expansion rate $\lambda$ defined as before,
\begin{equation}
\xi_{phys}\propto t^{1-\delta(1-\lambda)}\,.
\end{equation}

Neglecting the effect of the wall energy density on the background (specifically, on the Friedmann equations)---the relevant case for our numerical simulations---one can show that the attractor solution to the evolution equations (\ref{rhoevoldw1},\ref{vevoldw1}) is a linear scaling solution
\begin{equation}
L=\epsilon t\,, \qquad v=const\,.
\label{defscaling}
\end{equation}
Assuming that the scale factor behaves as $a \propto t^\lambda$,  the linear scaling constants above take the following detailed form:
\begin{equation}
\epsilon^2=\frac{k_w(k_w+c_w)}{3 \lambda (1-\lambda)}\,
\label{scaling1}
\end{equation}
\begin{equation}
v^2=\frac{1-\lambda}{3\lambda}\frac{k_w}{k_w+c_w}\,.
\label{scaling2}
\end{equation}

Numerically, we look for the best fit to the power laws
\begin{equation}
\frac{A}{V}\propto\rho_{w}\propto\frac{1}{\xi_{c}}\propto\eta^{\mu}\,,
\label{fit1}
\end{equation}
\begin{equation}
\gamma v\propto\eta^{\nu}\,;
\label{fit2}
\end{equation}
for a scale-invariant behavior, we should have $\mu=-1$ and $\nu=0$. The dynamics at the beginning of the simulation will be dominated by the initial conditions, while for the walls to be sufficiently well defined (which is important for accurately measuring walls areas and velocities) the co-moving correlation length should be significantly larger than the wall thickness. Our choice of the reliable period for the fits is done by inspection of each set of simulations, using these criteria \cite{Press}. Since we end all the simulations when the horizon becomes half the box size, the periodic boundary conditions should have no influence on our results.

In addition to measuring the scaling exponents $\mu$ and $\nu$, we will also be interested in the asymptotic values of $\xi_c/\eta$ and $\gamma v$, which can be related to the macroscopic parameters of the analytic model. These are calculated from the last few timesteps of each simulation, on the assumption that by then the network has reached scaling---note that our measured scaling exponents are consistent with this assumption. These are then used to calibrate the model. The scaling solution is parametrized by the expansion rate $\lambda$ (such that $a\propto t^\lambda$) and the phenomenological parameters $c_w$ and $k_w$. Given these parameters, the predicted values for  $\epsilon$ and $v$ are given by Eqs. (\ref{scaling1}--\ref{scaling2}), and therefore we can trivially obtain the value of $v$, while $\epsilon$ is given by
\begin{equation}
\epsilon=\frac{L}{t}=\frac{\xi_c}{(1-\lambda)\eta}\,;
\label{findc}
\end{equation}
from these one finally obtains the numerically measured values of $c_w$ and $k_w$. 

\section{Simulation results and model calibration}
\label{num}

\begin{figure}
\begin{center}
\includegraphics[width=2.6in]{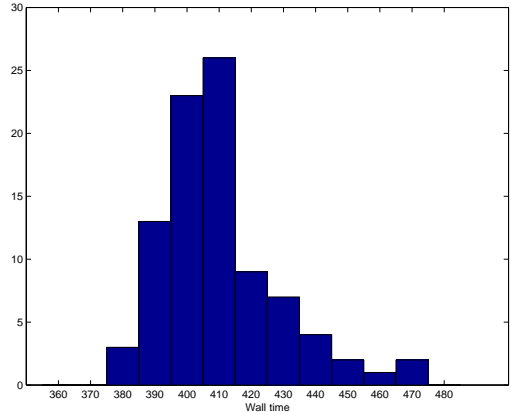}
\end{center}
\caption{\label{fig2}The distribution of the clock time needed to run each of our 90 $2048^3$ simulations on COSMOS, using 256 CPUs. The time of each run has been rounded off to the nearest ten-minute bin.}
\end{figure}

We have carried out three series of $2048^3$ simulations in the radiation era (corresponding to $\lambda=1/2$), the matter era ($\lambda=2/3$) and a fast-expansion era with $\lambda=4/5$. Each series consisted of 30 different simulations with different (random) initial conditions. These were run on the COSMOS supercomputer, using 256 CPU, and each took about 7 hours of clock time. The distribution of these times is shown in Fig. \ref{fig2}. The distribution of simulation times is mainly the result of the fact that the CPUs involved may not be physically close together on the machine: a simulation will start when (among other things) any 256 CPUs are available, and the closer they are to each other the faster the run time. (The time spent in identifying the walls can also vary from run to run. but we believe that this is a smaller effect.)

\begin{figure}
\begin{center}
\includegraphics[width=2.5in]{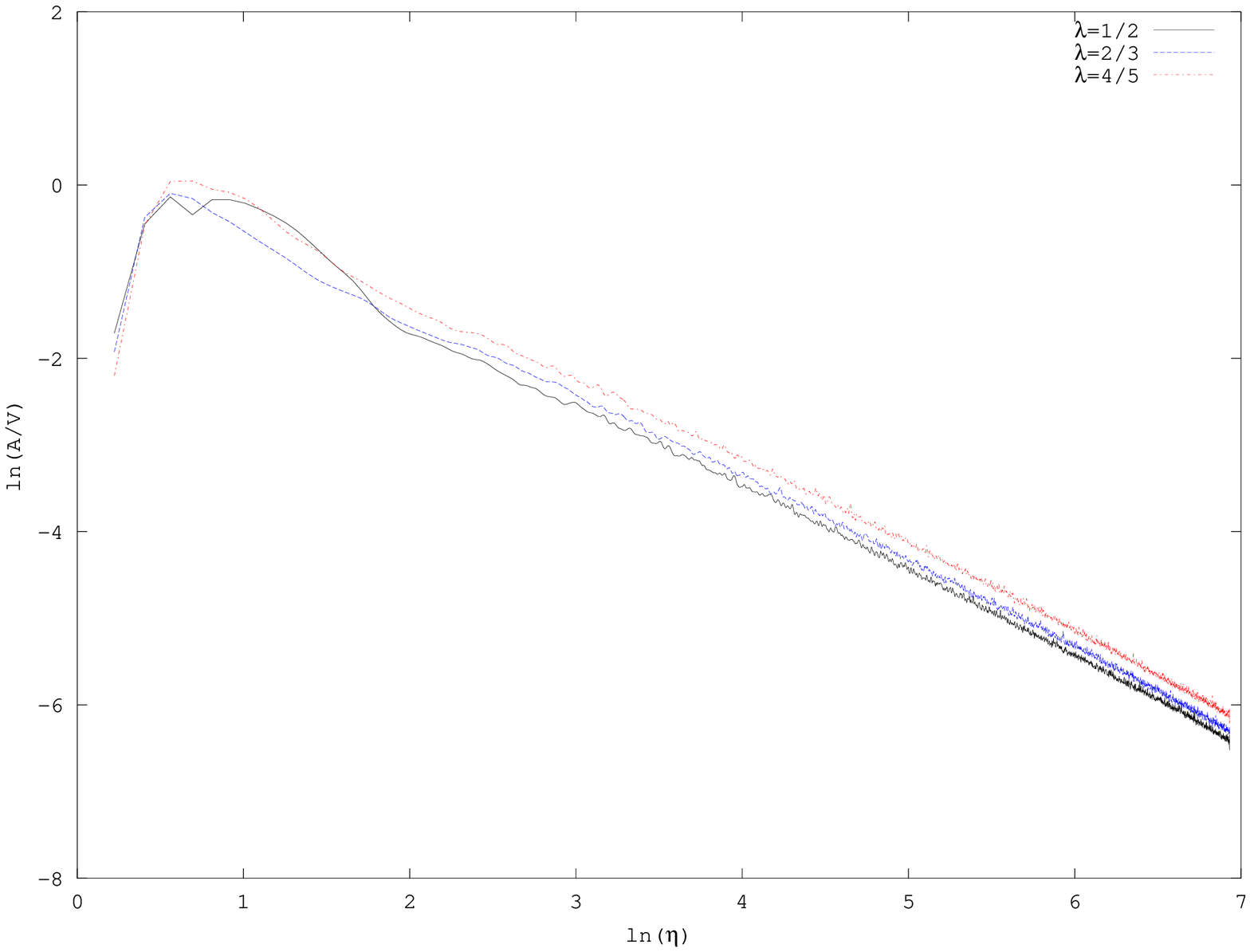}
\includegraphics[width=2.5in]{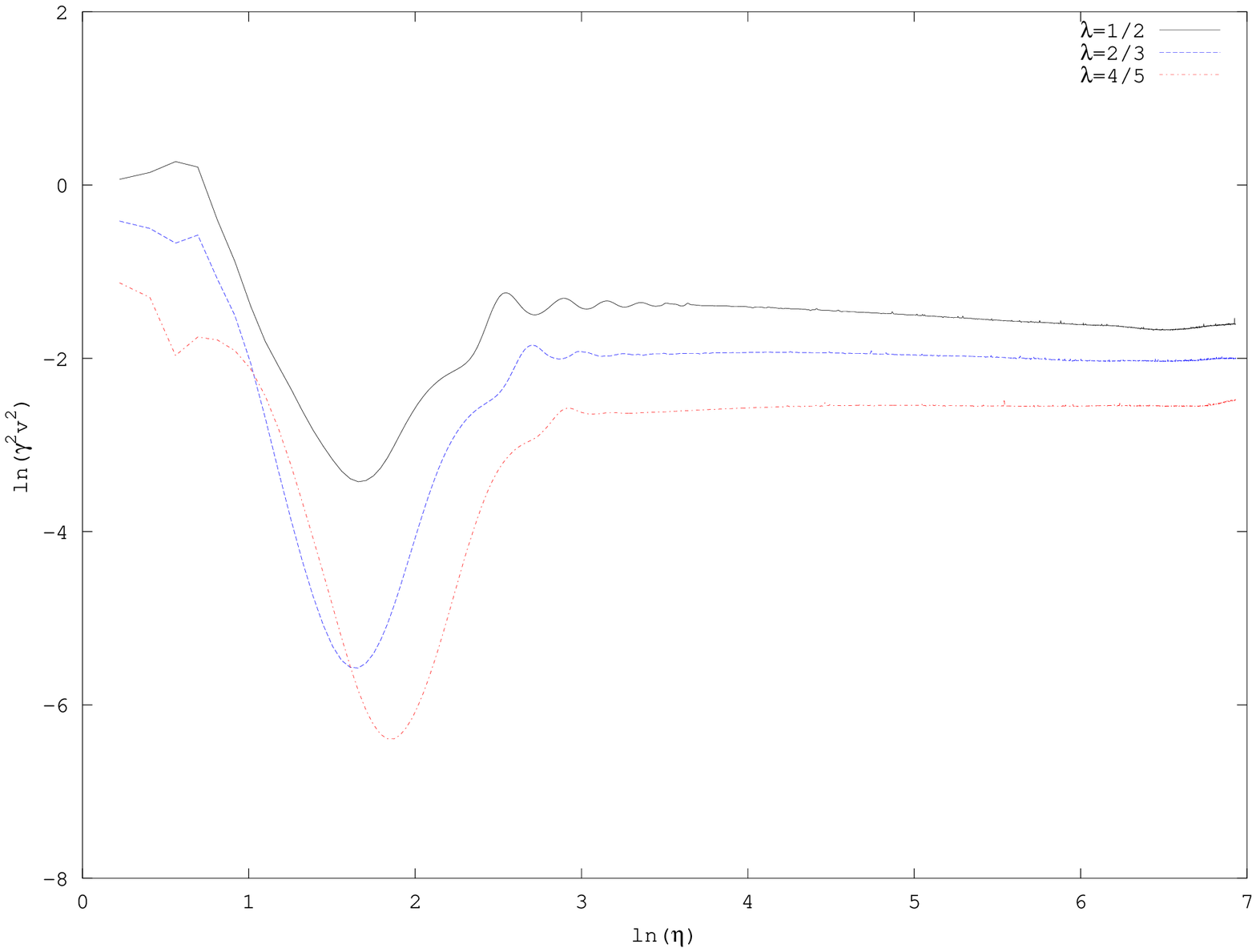}
\end{center}
\caption{\label{fig3}The evolution of the domain wall density ($\rho_w=A/V$) and velocity ($(\gamma v)^2$) as a function of conformal time, for the radiation era ($\lambda=1/2$), the matter era ($\lambda=2/3$) and the fast expansion era ($\lambda=4/5$). The plotted curves are the average of 30 different simulations with different (random) initial conditions.}
\end{figure}

Table \ref{tab1} and Fig. \ref{fig3} show the compared results of our series of runs. As can be seen from the values of the exponents $\mu$ and $\nu$, we find no deviation from the theoretically expected scaling solution. The behavior of the scaling properties in each of the three epochs is also to be expected. Smaller Hubble damping (corresponding to a smaller value of $\lambda$) allows the walls to have higher velocities. Therefore they will have more interactions in a given time, which leads to more energy losses, and consequently a smaller density (or equivalently a larger correlation length).

The oscillations in the wall velocities ar early times in Fig. \ref{fig3} are also worthy of notice. These arise as the network is relaxing away from the initial conditions and approaching the scaling solution. The fact that is is clearly visible here is in part due to the larger volume of the simulations (which provides better statistics), but mostly to the fact that we do not resort to an artificial damping period to make this relaxation faster.

\begin{table*}
\begin{center}
\begin{tabular}{|c|c|c|c|c|c|}
\hline 
$\lambda$&
Fit range&
$\mu$&
$\nu$&
$\xi_c/\eta$&
$\gamma v$\tabularnewline
\hline 
\hline 1/2 & 126-626 & $-1.00\pm0.04$ & $-0.00007\pm0.00004$ & $0.57\pm0.01$ & $0.45\pm0.05$ \tabularnewline
\hline 2/3 & 31-626 & $-0.99\pm0.02$ & $-0.00003\pm0.00002$ & $0.51\pm0.01$ & $0.37\pm0.03$ \tabularnewline
\hline 4/5 & 63.5-751 & $-1.01\pm0.03$ & $-0.00000\pm0.00001$ & $0.42\pm0.01$ & $0.29\pm0.02$ \tabularnewline
\hline
\end{tabular}
\end{center}
\caption{\label{tab1}The measured scaling properties of $2048^3$ PRS numerical simulations of domain wall networks with different expansion rates (parametrized by $\lambda$). The scaling exponents $\mu$ and $\nu$ were fitted using only the latter part of each simulation; the fit range (in conformal time) used for each set is shown in the second column. (See \protect\cite{Leite} for a discussion of the choice of this range.) The last two columns show the directly measured asymptotic values of $\xi_c/\eta$ and $\gamma v$, which can be related to the macroscopic parameters of the analytic model. All error bars are one-sigma statistical uncertainties obtained by averaging over 30 simulations with different random initial conditions.}
\end{table*}

\begin{figure}
\begin{center}
\includegraphics[width=2.5in,angle=0]{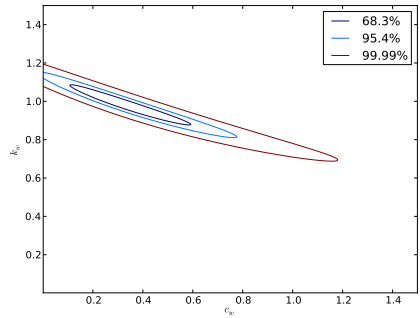}
\end{center}
\caption{\label{fig4}The likelihood contours for the VOS model parameters $c_w$ and $k_w$, using the directly measured asymptotic values of $\xi_c/\eta$ and $\gamma v$ shown in Table \protect\ref{tab1}.}
\end{figure}

The measured values of $\xi_c/\eta$ and $\gamma v$ in the three epochs can now be jointly used to calibrate the VOS model. Through a standard chi-squared minimization procedure we can easily obtain the likelihood contours in the $c_w-k_w$ plane shown in Fig. \ref{fig4}, and through marginalization we then obtain our final calibrated model parameters
\begin{equation}
c_w=0.34\pm0.16\,
\label{newcw}
\end{equation}
and
\begin{equation}
k_w=0.98\pm0.07\,;
\label{newkw}
\end{equation}
these results are remarkably consistent with an early qualitative analysis in \cite{MY2}, which had found $c_w\sim0.5$ and $k_w\sim0.9$, and with our own estimate in \cite{Leite}, which used sets of $1024^3$ runs and found $c_w=0.5\pm0.2$ and $k_w=1.1\pm0.3$.

The fact that the error bars on the model parameters are somewhat larger than those for $\xi_c/\eta$ and $\gamma v$ are an indication of the fact that taking $c_w$ and $k_w$ as constants is only an approximation, though a reasonable one in the context of a one-scale model. In other words, doing the fit separately for the three epochs would lead to slightly different best-fit parameters---cf. Table IV in \cite{Leite}. In particular, we might expect that $k_w$ is velocity-dependent, as was found to be the case for cosmic strings \cite{extend}.

\section{Conclusions}
\label{conc}

We have carried out the largest field theory simulations of standard domain wall networks undertaken to date, with simulation boxes of size $2048^3$, and confirmed that a scale-invariant evolution of the network, with $\xi_{phys}\propto t$ (or equivalently $\xi_c\propto \eta$) and $v=const$, is the attractor solution.

By comparing the network's asymptotic scaling properties in several cosmological epochs, we obtained an accurate calibration for the two free parameters in the velocity-dependent one-scale model for domain walls, one of which describes the network's energy loss rate while the other describes the (curvature-related) forces acting on the walls. Our result for these are respectively $c_w=0.34\pm0.16$ and $k_w=0.98\pm0.07$, which are significantly more precise than earlier estimates.

This combination of analytical and numerical techniques has also been used for studying cosmic strings \cite{Moore,Unified} and semilocal strings \cite{Nunes}, although in both cases the model calibration relies on significantly smaller simulations than the ones in the present paper. The larger number of degrees of freedom (corresponding to additional scalar fields, although for cosmic strings one has the alternative of Goto-Nambu simulations) makes the study of these models computationally challenging since in most cases the factor limiting the size of the boxes that can be simulated is memory rather than time, but otherwise our methods are directly applicable there. 

As previously said, domain walls are merely the simplest defect one can simulate. There remain certain types of well-motivated string (or hybrid) objects for which their physics, and most notably their cosmological evolution, remains relatively unexplored. An example is the evolution of cosmic strings carrying currents \cite{Oliveira}, even though such strings have been predicted in situations where more than one cosmological phase transition is considered, and they also arise naturally in SUSY models that predict strings. String currents also appear in the effective 4D description of higher dimensional strings, as the string position in the internal dimensions is described by worldsheet scalar fields giving rise to currents. Finally, there are superstrings models (or toy models thereof) where there are junctions connecting string segments which have different masses per unit length and different intercommuting properties.

All of these more realistic models have more degrees of freedom than domain walls, and simulating them therefore requires, for a given box size, an amount of memory that increases at least as the number of degrees of freedom. Additionally, the extra complexity also implies that each evolution timestep will take longer in these models. Nevertheless, available codes seem to scale reasonably well, at least in the range of conditions probed so far. There is therefore a case for carrying out simulations using ca. 1000 cpus and 1Tb of memory.

\section*{Acknowledgments}
The work of CM is funded by a Ci\^encia2007 Research Contract, funded by FCT/MCTES (Portugal) and POPH/FSE (EC), and we acknowledge additional support from project PTDC/FIS/111725/2009 from FCT, Portugal.

The numerical simulations in this paper were performed on the COSMOS Consortium supercomputer within the DiRAC Facility jointly funded by STFC and the Large Facilities Capital Fund of BIS (UK). We are grateful to Andrey Kaliazin for his help with the code optimization.

\bibliographystyle{model1-num-names}
\bibliography{walls}

\begin{thebibliography}{22}
\expandafter\ifx\csname natexlab\endcsname\relax\def\natexlab#1{#1}\fi
\providecommand{\bibinfo}[2]{#2}
\ifx\xfnm\relax \def\xfnm[#1]{\unskip,\space#1}\fi
%Type = Article
\bibitem[{Kibble(1976)}]{Kibble}
\bibinfo{author}{T.~W.~B. Kibble},
\newblock \bibinfo{title}{Topology of cosmic domains and strings},
\newblock \bibinfo{journal}{J. Phys.} \bibinfo{volume}{A9}
  (\bibinfo{year}{1976}) \bibinfo{pages}{1387--1398}.
%Type = Book
\bibitem[{Vilenkin and Shellard(1994)}]{Book}
\bibinfo{author}{A.~Vilenkin}, \bibinfo{author}{E.~P.~S. Shellard},
  \bibinfo{title}{COSMIC STRINGS AND OTHER TOPOLOGICAL DEFECTS},
  \bibinfo{year}{1994}. \bibinfo{note}{{ }Cambridge, U.K.: Cambridge University
  Press}.
%Type = Article
\bibitem[{Zeldovich et~al.(1974)Zeldovich, Kobzarev, and Okun}]{Zeldovich}
\bibinfo{author}{Y.~B. Zeldovich}, \bibinfo{author}{I.~Y. Kobzarev},
  \bibinfo{author}{L.~B. Okun},
\newblock \bibinfo{title}{Cosmological consequences of the spontaneous
  breakdown of discrete symmetry},
\newblock \bibinfo{journal}{Zh. Eksp. Teor. Fiz.} \bibinfo{volume}{67}
  (\bibinfo{year}{1974}) \bibinfo{pages}{3--11}.
%Type = Article
\bibitem[{Avelino et~al.(2008)Avelino, Martins, Menezes, Menezes, and
  Oliveira}]{Junctions}
\bibinfo{author}{P.~P. Avelino}, \bibinfo{author}{C.~J. A.~P. Martins},
  \bibinfo{author}{J.~Menezes}, \bibinfo{author}{R.~Menezes},
  \bibinfo{author}{J.~C. R.~E. Oliveira},
\newblock \bibinfo{title}{{Dynamics of domain wall networks with junctions}},
\newblock \bibinfo{journal}{Phys.Rev.} \bibinfo{volume}{D78}
  (\bibinfo{year}{2008}) \bibinfo{pages}{103508}.
%Type = Article
\bibitem[{Press et~al.(1989)Press, Ryden, and Spergel}]{Press}
\bibinfo{author}{W.~H. Press}, \bibinfo{author}{B.~S. Ryden},
  \bibinfo{author}{D.~N. Spergel},
\newblock \bibinfo{title}{Dynamical evolution of domain walls in an expanding
  universe},
\newblock \bibinfo{journal}{Astrophys. J.} \bibinfo{volume}{347}
  (\bibinfo{year}{1989}) \bibinfo{pages}{590}.
%Type = Article
\bibitem[{Leite and Martins(2011)}]{Leite}
\bibinfo{author}{A.~M.~M. Leite}, \bibinfo{author}{C.~J. A.~P. Martins},
\newblock \bibinfo{title}{{Scaling Properties of Domain Wall Networks}},
\newblock \bibinfo{journal}{Phys.Rev.} \bibinfo{volume}{D84}
  (\bibinfo{year}{2011}) \bibinfo{pages}{103523}.
%Type = Article
\bibitem[{Coulson et~al.(1996)Coulson, Lalak, and Ovrut}]{Coulson}
\bibinfo{author}{D.~Coulson}, \bibinfo{author}{Z.~Lalak},
  \bibinfo{author}{B.~A. Ovrut},
\newblock \bibinfo{title}{Biased domain walls},
\newblock \bibinfo{journal}{Phys. Rev.} \bibinfo{volume}{D53}
  (\bibinfo{year}{1996}) \bibinfo{pages}{4237--4246}.
%Type = Article
\bibitem[{Larsson et~al.(1997)Larsson, Sarkar, and White}]{Larsson}
\bibinfo{author}{S.~E. Larsson}, \bibinfo{author}{S.~Sarkar},
  \bibinfo{author}{P.~L. White},
\newblock \bibinfo{title}{Evading the cosmological domain wall problem},
\newblock \bibinfo{journal}{Phys. Rev.} \bibinfo{volume}{D55}
  (\bibinfo{year}{1997}) \bibinfo{pages}{5129--5135}.
%Type = Article
\bibitem[{Avelino and Martins(2000)}]{Fossils}
\bibinfo{author}{P.~P. Avelino}, \bibinfo{author}{C.~J. A.~P. Martins},
\newblock \bibinfo{title}{Topological defects: Fossils of an anisotropic era?},
\newblock \bibinfo{journal}{Phys. Rev.} \bibinfo{volume}{D62}
  (\bibinfo{year}{2000}) \bibinfo{pages}{103510}.
%Type = Article
\bibitem[{Garagounis and Hindmarsh(2003)}]{Garagounis}
\bibinfo{author}{T.~Garagounis}, \bibinfo{author}{M.~Hindmarsh},
\newblock \bibinfo{title}{Scaling in numerical simulations of domain walls},
\newblock \bibinfo{journal}{Phys. Rev.} \bibinfo{volume}{D68}
  (\bibinfo{year}{2003}) \bibinfo{pages}{103506}.
%Type = Article
\bibitem[{Oliveira et~al.(2005)Oliveira, Martins, and Avelino}]{MY1}
\bibinfo{author}{J.~C. R.~E. Oliveira}, \bibinfo{author}{C.~J. A.~P. Martins},
  \bibinfo{author}{P.~P. Avelino},
\newblock \bibinfo{title}{{The Cosmological evolution of domain wall
  networks}},
\newblock \bibinfo{journal}{Phys.Rev.} \bibinfo{volume}{D71}
  (\bibinfo{year}{2005}) \bibinfo{pages}{083509}.
%Type = Article
\bibitem[{Avelino et~al.(2005)Avelino, Martins, and Oliveira}]{MY2}
\bibinfo{author}{P.~P. Avelino}, \bibinfo{author}{C.~J. A.~P. Martins},
  \bibinfo{author}{J.~C. R.~E. Oliveira},
\newblock \bibinfo{title}{{One-scale model for domain wall network evolution}},
\newblock \bibinfo{journal}{Phys.Rev.} \bibinfo{volume}{D72}
  (\bibinfo{year}{2005}) \bibinfo{pages}{083506}.
%Type = Article
\bibitem[{Lalak et~al.(2008)Lalak, Lola, and Magnowski}]{Lalak}
\bibinfo{author}{Z.~Lalak}, \bibinfo{author}{S.~Lola},
  \bibinfo{author}{P.~Magnowski},
\newblock \bibinfo{title}{{Dynamics of domain walls for split and runaway
  potentials}},
\newblock \bibinfo{journal}{Phys.Rev.} \bibinfo{volume}{D78}
  (\bibinfo{year}{2008}) \bibinfo{pages}{085020}.
%Type = Article
\bibitem[{Hindmarsh(2003)}]{Hindmarsh}
\bibinfo{author}{M.~Hindmarsh},
\newblock \bibinfo{title}{Level set method for the evolution of defect and
  brane networks},
\newblock \bibinfo{journal}{Phys. Rev.} \bibinfo{volume}{D68}
  (\bibinfo{year}{2003}) \bibinfo{pages}{043510}.
%Type = Article
\bibitem[{Martins and Shellard(1996{\natexlab{a}})}]{ms1a}
\bibinfo{author}{C.~J. A.~P. Martins}, \bibinfo{author}{E.~P.~S. Shellard},
\newblock \bibinfo{title}{String evolution with friction},
\newblock \bibinfo{journal}{Phys. Rev.} \bibinfo{volume}{D53}
  (\bibinfo{year}{1996}{\natexlab{a}}) \bibinfo{pages}{575--579}.
%Type = Article
\bibitem[{Martins and Shellard(1996{\natexlab{b}})}]{ms1b}
\bibinfo{author}{C.~J. A.~P. Martins}, \bibinfo{author}{E.~P.~S. Shellard},
\newblock \bibinfo{title}{Quantitative string evolution},
\newblock \bibinfo{journal}{Phys. Rev.} \bibinfo{volume}{D54}
  (\bibinfo{year}{1996}{\natexlab{b}}) \bibinfo{pages}{2535--2556}.
%Type = Article
\bibitem[{Martins and Shellard(2002)}]{extend}
\bibinfo{author}{C.~J. A.~P. Martins}, \bibinfo{author}{E.~P.~S. Shellard},
\newblock \bibinfo{title}{Extending the velocity-dependent one-scale string
  evolution model},
\newblock \bibinfo{journal}{Phys. Rev.} \bibinfo{volume}{D65}
  (\bibinfo{year}{2002}) \bibinfo{pages}{043514}.
%Type = Article
\bibitem[{Moore et~al.(2002)Moore, Shellard, and Martins}]{Moore}
\bibinfo{author}{J.~N. Moore}, \bibinfo{author}{E.~P.~S. Shellard},
  \bibinfo{author}{C.~J. A.~P. Martins},
\newblock \bibinfo{title}{On the evolution of abelian-higgs string networks},
\newblock \bibinfo{journal}{Phys. Rev.} \bibinfo{volume}{D65}
  (\bibinfo{year}{2002}) \bibinfo{pages}{023503}.
%Type = Article
\bibitem[{Martins and Achucarro(2008)}]{Monopoles}
\bibinfo{author}{C.~J. A.~P. Martins}, \bibinfo{author}{A.~Achucarro},
\newblock \bibinfo{title}{{Evolution of local and global monopole networks}},
\newblock \bibinfo{journal}{Phys.Rev.} \bibinfo{volume}{D78}
  (\bibinfo{year}{2008}) \bibinfo{pages}{083541}.
%Type = Article
\bibitem[{Martins et~al.(2004)Martins, Moore, and Shellard}]{Unified}
\bibinfo{author}{C.~J. A.~P. Martins}, \bibinfo{author}{J.~N. Moore},
  \bibinfo{author}{E.~P.~S. Shellard},
\newblock \bibinfo{title}{{A Unified model for vortex string network
  evolution}},
\newblock \bibinfo{journal}{Phys.Rev.Lett.} \bibinfo{volume}{92}
  (\bibinfo{year}{2004}) \bibinfo{pages}{251601}.
%Type = Article
\bibitem[{Nunes et~al.(2011)Nunes, Avgoustidis, Martins, and
  Urrestilla}]{Nunes}
\bibinfo{author}{A.~S. Nunes}, \bibinfo{author}{A.~Avgoustidis},
  \bibinfo{author}{C.~J. A.~P. Martins}, \bibinfo{author}{J.~Urrestilla},
\newblock \bibinfo{title}{{Analytic Models for the Evolution of Semilocal
  String Networks}},
\newblock \bibinfo{journal}{Phys.Rev.} \bibinfo{volume}{D84}
  (\bibinfo{year}{2011}) \bibinfo{pages}{063504}.
%Type = Article
\bibitem[{Oliveira et~al.(2012)Oliveira, Avgoustidis, and Martins}]{Oliveira}
\bibinfo{author}{M.~F. Oliveira}, \bibinfo{author}{A.~Avgoustidis},
  \bibinfo{author}{C.~J. A.~P. Martins},
\newblock \bibinfo{title}{{Cosmic string evolution with a conserved charge}},
\newblock \bibinfo{journal}{Phys.Rev.} \bibinfo{volume}{D85}
  (\bibinfo{year}{2012}) \bibinfo{pages}{083515}.

\end{thebibliography}

\end{document}